\documentclass[lettersize,journal]{IEEEtran}
\usepackage{amsmath,amsfonts}
\usepackage{algorithmic}
\usepackage{array}
\usepackage{textcomp}
\usepackage{stfloats}
\usepackage{url}
\usepackage{verbatim}
\usepackage{graphicx}
\usepackage{subfigure} 
\usepackage{amsthm}
\usepackage[hidelinks]{hyperref}            % For hyperlinks without rounding box
\def\BibTeX{{\rm B\kern-.05em{\sc i\kern-.025em b}\kern-.08em
		T\kern-.1667em\lower.7ex\hbox{E}\kern-.125emX}}
\usepackage{balance}
\usepackage{makecell}
\newcommand{\re}[1]{(\ref{#1})}
\newtheorem{remark}{Remark}
\newtheorem{assumption}{Assumption}

\hypersetup{
	colorlinks=true,
	linkcolor=green,
	citecolor=blue
}

\begin{document}
	\title{Safe Reinforcement Learning-Based Eco-Driving Control for Mixed Traffic Flows With Disturbances
	}
	\author{Ke Lu, Dongjun Li, Qun Wang, Kaidi Yang, Lin Zhao, and Ziyou Song
		\thanks{Ke Lu is with the Department of Mechanical Engineering, National University of Singapore, Singapore 117575, Singapore, and also with the School of Automation Science and Engineering, South China University of Technology, Guangzhou 510641, China (e-mail: lukenanlin@163.com).}
		\thanks{Dongjun Li, Qun Wang, and Ziyou Song are with the Department of Mechanical Engineering, National University of Singapore, Singapore 117575, Singapore (e-mail: dongjun.li@u.nus.edu; wangqun0516@gmail.com; ziyou@nus.edu.sg).}
		\thanks{Kaidi Yang is with the Department of Civil and Environmental Engineering, National University of Singapore, Singapore 117576, Singapore (e-mail: ykaidi@nus.edu.sg).}
		\thanks{Lin Zhao is with the Department of Electrical and Computer Engineering, National University of Singapore, Singapore 119077, Singapore (e-mail: elezhli@nus.edu.sg).}
		\thanks{Corresponding author: Ziyou Song.}
	}

	\markboth{}%IEEE Transactions on Transportation Electrification
	{How to Use the IEEEtran \LaTeX \ Templates}
	
	\maketitle
	
	\begin{abstract}
		This paper presents a safe learning-based eco-driving framework tailored for mixed traffic flows, which aims to optimize energy efficiency while guaranteeing safety during real-system operations.
		Even though reinforcement learning (RL) is capable of optimizing energy efficiency in intricate environments, it is challenged by safety requirements during the training process. The lack of safety guarantees is the other concern when deploying a trained policy in real-world application.
		Compared with RL, model predicted control (MPC) can handle constrained dynamics systems, ensuring safe driving. However, the major challenges lie in complicated eco-driving tasks and the presence of disturbances, which respectively challenge the MPC design and the satisfaction of constraints.
		To address these limitations, the proposed framework incorporates the tube-based enhanced MPC (RMPC) to ensure the safe execution of the RL policy under disturbances, thereby improving the control robustness.
		RL not only optimizes the energy efficiency of the connected and automated vehicle in mixed traffic but also handles more uncertain scenarios, in which the energy consumption of the human-driven vehicle and its diverse and stochastic driving behaviors are considered in the optimization framework.
		Simulation results demonstrate that the proposed algorithm, compared with RMPC technique, shows an average improvement of $10.88\%$ in holistic energy efficiency, while compared with RL algorithm, it effectively prevents inter-vehicle collisions.
	\end{abstract}
	
	\begin{IEEEkeywords}
		Eco-driving, safe reinforcement learning, robust model predictive control, mixed traffic, connected and automated vehicles.
	\end{IEEEkeywords}

	\section{Introduction}
	\IEEEPARstart{E}{nergy} consumption and pollutant emissions have become urgent global concerns, driving efforts to improve vehicle energy efficiency \cite{Shao2021VehicleSpeed,Yang2021EcoDriving}. In particular, the eco-driving strategy, which commonly refers to optimizing driving styles to reduce energy consumption, has received great attention \cite{Dong2023Flexible,ZHANG2021116215}. With advancements in autonomous driving and Vehicle-to-Everything technologies, real-time access to traffic information and road preview allows for the precise control and trajectory planning of connected and automated vehicles (CAVs), which provides a promising approach to eco-driving strategies \cite{VAHIDI2018822,GUANETTI201818}. Consequently, many CAV-based techniques have been developed to improve energy efficiency in various traffic scenarios such as highway \cite{dong2022predictive}  and signalized intersection \cite{Meng2022EcoDriving}.
	Compared with a pure CAV flow, in general, there is a consensus that CAVs will coexist with human-driven vehicles (HDVs) in traffic flow for a long period, which promotes efforts to reduce energy consumption in mixed traffic. 
	For example, in \cite{ZHAO2018802}, a real-time cooperative eco-driving strategy for mixed CAVs and HDVs approaching a signalized intersection is proposed to minimize the total fuel consumption of the whole platoon.
	A Data-EnablEd Predictive Leading Cruise Control (DeeP-LCC) strategy is presented in \cite{Wang2023DeePLCC}, which improves driving safety, fuel economy, and traffic smoothness in mixed traffic flow. 
	In \cite{Hu2021EcoPlatooning}, the leading vehicle of the platoon is responsible for velocity optimization to reduce energy consumption while interacting with the preceding HDV. 
	
	Reinforcement learning (RL) is a powerful technique for learning an optimal control policy in complex environments, which has been widely applied to the control of CAVs for eco-driving. In \cite{WANG2023120563}, a multi-agent RL approach is proposed to deal with the cooperative optimization of eco-driving and energy management, which minimizes energy consumption while simultaneously maintaining a safe following distance. 
	A novel RL-based framework for optimizing fuel economy and driving safety is presented in \cite{Liu2020Enhancing}, where autonomous vehicles attain a collision-free driving policy by learning from mistakes. 
	In \cite{Bai2022Hybrid}, a hybrid RL framework is developed to learn complex driving strategies from conflicting factors such as speeding up against energy-saving at signalized intersections.
	Even though RL is capable of learning a safe policy from trial and error, e.g., acquiring a collision-free driving policy from repeated collisions, various safety hazards exist in practice and can lead to accidents during the learning process. 
	Furthermore, there is no guarantee that the trained policy can avoid collisions when applied to real systems.
	Therefore, how to safely learn an eco-driving policy and guarantee the safe driving of the trained policy at all times is one of the motivations for this paper.
	
	In practical engineering systems, it is essential that the system constraints are never violated during operation \cite{Jin2019Adaptive,Lv2022Finite}.
	Model predictive control (MPC) is an optimization-based technique \cite{MAYNE20142967,QIN2003733}, which has been widely utilized to address constraint requirements in various eco-driving strategies. 
	For example, a safety-enhanced eco-driving strategy for CAVs is proposed in \cite{ZHOU2023104320}, where MPC is integrated into the optimization procedure to reduce the driving risk. 
	In \cite{Moser2018Flexible}, a stochastic MPC approach is presented to optimize fuel consumption while simultaneously dropping the number of constraint violations in a car-following context. 
	A hierarchical velocity optimization design based on hybrid MPC is developed in \cite{Zhang2023Hierarchical} to reduce fuel consumption and pollution emission while guaranteeing the safety velocity constraint.
	Furthermore, as a practical problem, the presence of disturbances may result in the violations of system constraints, which promotes the development of robust MPC (RMPC) technique \cite{MAYNE2005219,Mayne2011Tube}. 
	Consequently, various RMPC-based algorithms are employed to handle the bounded disturbances and ensure constraint requirements in cooperative adaptive cruise control \cite{Feng2020TubeBased} and vehicle platooning control \cite{Feng2021RobustPlatoon,Zhou2022Robust}.
	However, optimizing the nonlinear energy consumption in complex environments, not limited to disturbances, while guaranteeing constraint requirements, challenges the RMPC design.
	
	The combination of RL and MPC aims to exploit the benefits of both fields \cite{Koller2018Learning,Zanon2021Safe,Wabersich2018Linear,WABERSICH2021109597}. Inspired by \cite{Wabersich2018Linear,WABERSICH2021109597}, a safe learning-based eco-driving framework for mixed traffic flows is developed. Compared with RL-based methods, the incorporated RMPC can ensure the safe execution of the RL policy under disturbances at all times. Compared with MPC-based methods, the integrated RL can optimize energy efficiency in complex environments, which reduces the complexity of RMPC design.
	In addition to optimizing the energy consumption of the CAV, RL in the presented framework can also be fully utilized to handle more complicated eco-driving issues, e.g., the energy consumption of coexisted HDV and its diverse and stochastic driving behaviors.
	In the context of eco-driving, the car-following models such as optimal velocity model (OVM) \cite{bando1995dynamical} and intelligent driver model (IDM) \cite{treiber2000congested} are utilized to formulate the dynamics of HDVs in mixed traffic flow, such that HDVs can be further considered in the optimization framework \cite{CHEN2021103138,Yu2022AnEcoDriving,Jiang2023Learning}. Recently, a fuel-economical driving strategy for CAVs, along with its fuel-economy impacts on a human-driven platoon is first studied in \cite{Ma2022Energetic}, where individual drivers' behavior variations are considered.
	Therefore, based on the presented framework, a more complex eco-driving problem is investigated, where not only the holistic energy consumption but also the diverse and stochastic driving behaviors of the following HDV are incorporated into the optimization procedure.
	
	This paper presents a safe learning-based eco-driving framework for mixed traffic flows, which aims to optimize energy efficiency while guaranteeing safety.
	RL can be fully utilized to address complicated eco-driving problems, e.g., optimizing energy efficiency while considering drivers' behavior variations.
	Accordingly, the impacts of diverse and stochastic behaviors of the following HDV on holistic energy consumption are investigated.
        To overcome the safety concern arising from RL, the incorporated MPC ensures the safe learning and deployment of complex eco-driving policies in real-world applications.
        In addition, as a practical problem, the disturbances, which stem from measurement noise and prediction uncertainty, impose a great challenge on the constrained control.
	To improve the robustness, tube-based enhanced MPC is introduced to guarantee system constraints such as safety under disturbances.
	The main contributions of this work are summarized as follows.
	\begin{itemize}
		\item[(i)] A safe learning-based eco-driving strategy for mixed traffic flows is proposed, thereby enabling the safe learning and deployment of complex eco-driving policies.
		
		\item[(ii)] Tube-based enhanced MPC is introduced to improve the control robustness, ensuring the satisfaction of system constraints despite the presence of disturbances.
		
		\item[(iii)] The energy consumption of both the CAV and the HDV are integrated into the presented framework, in which the impacts of diverse and stochastic driving behaviors of HDV on eco-driving performance are investigated.
		
	\end{itemize}
	
	\section{Problem Formulation}\label{Formulation}
	
	\subsection{Notations and Scenario}
	
	The Minkowski sum of two sets is $\mathbb{A} \oplus \mathbb{B} := \{a+b| a\in \mathbb{A}, b \in \mathbb{B} \}$. The Pontryagin set difference is $\mathbb{A} \ominus \mathbb{B} := \{a| a + b \in \mathbb{A}, b \in \mathbb{B}\}$. 
	A set $\mathbb{Z}$ is robustly positive invariant for the discrete-time system $x(k+1) = f(x(k),w(k))$ where $w(k) \in \mathbb{W}$ if $f(x(k),w(k)) \in \mathbb{Z}$ for all $x(k) \in \mathbb{Z}$ and $w(k) \in \mathbb{W}$.
	
\begin{figure}[!thb]
		\centering
		\includegraphics[width=1\linewidth]{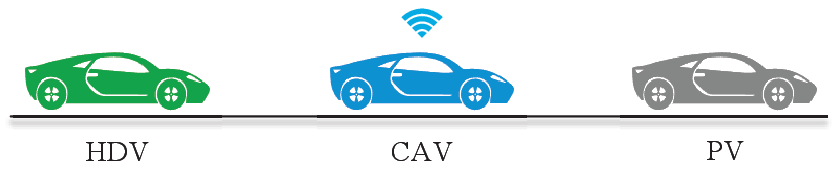}
		\caption{Illustration of mixed traffic flow.}
		\label{fig_Scenario}
	\end{figure}

	A simple scenario of mixed traffic is considered in this paper, where a human-driven preceding vehicle (PV), a CAV, and the following HDV share the same lane, as shown in Fig. \ref{fig_Scenario}. The CAV can measure the distances and velocities of both the PV and the following HDV via onboard sensors (e.g., millimeter-wave radars). There can exist more vehicles before the PV and behind the HDV, so the scenario is a fraction of a long mixed traffic flow.
	
	\subsection{System Model}
	Consider the following discrete dynamics of vehicles 
	\begin{equation}\label{vehicle_model}
		x_i(k+1) = A x_i(k) + B u_i(k), ~ i \in \{p, c, h\}
	\end{equation}
	with
	\begin{equation}
		x_i = \left[
		\begin{array}{c}
			s_i\\
			v_i\nonumber
		\end{array}
		\right],~  
		A =
		\left[
		\begin{array}{ccc}
			1& \tau  \\
			0& 1 \nonumber
		\end{array}
		\right],~
		B =
		\left[
		\begin{array}{c}
			0.5 \tau^2\\
			\tau \nonumber
		\end{array}
		\right] 
		\nonumber
	\end{equation}
	where $x_p$, $x_c$, and $x_h$ respectively denote the states of PV, CAV, and HDV, $s_i$ and $v_i$ denote the position and velocity of the vehicle, respectively, $u_c$ is the acceleration control input of CAV, $u_i = a_i$, $i \in \{p,h\}$ denote the accelerations of PV and HDV, respectively, $k$ denotes the discrete time step, and $\tau$ is the sampling time interval.
	Applying the constant time headway range policy and vehicle dynamics \re{vehicle_model}, the car-following system is formulated as
	\begin{equation}\label{carfollowing_system}
		x(k) = x_p(k) + H x_c(k)
	\end{equation}
	with
	\begin{equation}
		H =
		\left[
		\begin{array}{cc}
			-1& -h \\
			0& -1 \nonumber
		\end{array}
		\right]
		\nonumber
	\end{equation}
	where $h$ is the constant time headway.
	In practice, the CAV cannot precisely capture the dynamics of the PV due to the existence of disturbances such as prediction uncertainties and measurement errors in real environments. Thus, let the nominal systems of PV and CAV corresponding to \re{vehicle_model} be defined by
	\begin{equation}\label{nominal_vehicle}
		\bar x_i(k+1) = A \bar x_i(k) + B \bar u_i(k),~ i \in \{p,c\}
	\end{equation}
	where $\bar x_i$ and $\bar u_i$ are the planned state and acceleration, respectively, with $\bar u_p = \bar a_p$ being the predicted acceleration of the PV.
	In terms of \re{carfollowing_system}, the predicted state is derived as
	\begin{equation}\label{nominal_system}
		\bar x(k) = \bar x_p(k) + H \bar x_c(k).
	\end{equation}
        Substituting \re{nominal_vehicle} into \re{nominal_system}, the nominal car-following system is given by
        \begin{align}\label{dy_nom_carfollowing}
		\bar x(k+1) =&~ \bar x_p(k+1) + H \bar x_c(k+1)\nonumber\\
		            =&~ A \bar x(k) + B_c \bar u_c(k) + B \bar a_p(k)
	\end{align}
        where $HA = AH$, $B_c = HB$. Consider the noise in the measurement of PV, the uncertain system corresponding to PV given in \re{vehicle_model} is obtained
        \begin{equation}\label{disturbance_PV}
		x_p(k+1) = A x_p(k) + B a_p(k) + w_n(k)
	\end{equation}
        where $w_n(k)$ is the bounded disturbance resulted from the measurement noise.
	In addition, although there are various methods that can be employed for the prediction, the human-driven PV does not follow deterministic driving behaviors, which implies the existence of prediction uncertainty.
        Accordingly, denote the disturbance resulted from the prediction error as $w_p(k) = B(\bar a_p(k) - a_p(k))$, and thus it follows from \re{carfollowing_system} and \re{disturbance_PV} that the actual car-following system is derived as
        \begin{equation}\label{dy_act_carfollowing}
		x(k+1) = A x(k) + B_c u_c(k) + B \bar a_p(k) + w(k)
	\end{equation}
        where $w(k) = w_n(k) - w_p(k) $ is the disturbance resulted from both the measurement noise and the prediction error, and the disturbance is assumed to be bounded, that is, $w(k) \in \mathbb{W} = \{ w(k) \in \mathbb{R}^2: ||w(k)||_{\infty} \leq c_w \} $.
        Then, define the deviation between the actual and nominal systems corresponding to \re{carfollowing_system} as
	\begin{equation}\label{deviation_act_nom}
		\tilde{x}(k) = x(k) - \bar{x}(k).
	\end{equation}
	Substituting \re{dy_nom_carfollowing} and \re{dy_act_carfollowing} into \re{deviation_act_nom} yields
        \begin{align}\label{deviation_error}
		\tilde{x}(k+1) =&~ x(k+1) - \bar{x}(k+1)\nonumber\\
		               =&~ A \tilde{x}(k) + B_c [u_c(k) - \bar u_c(k)] + w(k).
	\end{align}
	Furthermore, to ensure the safe driving and physical limits, the state and control input of the actual system \re{dy_act_carfollowing} are subject to the following constraints 
	\begin{align}\label{System_constraints}
        \begin{split}
		x(k) \in \mathbb{X} =&~ \left\{
		\begin{aligned}
			&-x_{1,\min} \leq x_1(k) \\
			&-x_{2,\min} \leq x_2(k) \leq x_{2,\max}
		\end{aligned}
		\right\}\\
		u_c(k) \in \mathbb{U} =&~  \{ -u_{\min} \leq u_c(k) \leq u_{\max} \}
        \end{split}
	\end{align}
	where $-x_{1,\min}$ is the minimal distance for the safety requirement, $-x_{2,\min}$ and $ x_{2,\max}$ are the lower and upper bounds of the velocity difference, respectively, and $-u_{\min}$ and $u_{\max}$ denote the minimum and maximum accelerations.
	
	\begin{remark}
		In the car-following scenario, it is assumed that the human driver does not actively collide with its preceding CAV and can brake in time. Thus, the constraints between the CAV and the HDV are not considered in the control design.
	\end{remark}
	
	\subsection{Energy Consumption Model}
	
	The energy consumption of electric vehicles is specified as an integration of output power at the battery terminal, which consists of the propulsion and regenerative braking powers \cite{ehsani2018modern,Ye2016Ahybrid}. Thus, the instantaneous energy consumption model is given by
	\begin{equation}
		P_{\textrm{battery}} = P_{\textrm{b,out}} + P_{\textrm{b,in}}
		\nonumber
	\end{equation}
	with
	\begin{align}
		P_{\textrm{b,out}} =&~ \frac{v}{\eta_t \eta_m} P_{\textrm{b,var}}, \quad P_{\textrm{b,in}} = k_0 v \eta_t \eta_m P_{\textrm{b,var}}\nonumber\\
		P_{\textrm{b,var}} =&~ mg(f\cos\alpha + \sin\alpha) + 0.5\rho C_D A_f v^2 + m \delta \frac{dv}{dt}
		\nonumber
	\end{align}
	where $\eta_t$ and $\eta_m$ are commonly approximated by constant values, respectively denoting the transmission efficiency and motor drive efficiency, $m$ is the mass of the electric vehicle, $f$ is the rolling resistance coefficient, $g$ is the gravitational constant, $\rho$ is the air density, $C_D$ is aerodynamic drag coefficient, $A_f$ is the frontal area of electric vehicle, $\delta$ is the vehicle mass related coefficient, $\alpha$ is the road grade, and $k_0$ is the regenerative braking factor.
	To avoid heavy computation and complicated calibration, an approximated and differentiable energy consumption model in a polynomial expression is fitted \cite{wang2022adaptive} as follows
	\begin{equation}\label{Energy_model_poly}
		P_{\textrm{mot}}(v,a) = \sum_{i=0}^{3} \sum_{j=0}^{2} p_{ij} v^i(t) a^j(t)
	\end{equation}
	where $v(t)$ and $a(t)$ are instantaneous velocity and acceleration, respectively, and $p_{ij}$ are fitted parameters listed in Table \ref{Para_model_con}. Those fitted parameters not listed in Table \ref{Para_model_con} are set to zero.
	
	\begin{table}[!ht] 
		\centering 
		\caption{ENERGY CONSUMPTION MODEL PARAMETERS}
		\begin{tabular}{l|l|l|l|l|l} \hline 
			\multicolumn{1}{l}{Parameter} & Value & \multicolumn{1}{l}{Parameter} & Value & \multicolumn{1}{l}{Parameter} & Value \\ \hline
			\multicolumn{1}{l}{$p_{00}$} & $110.3$ & \multicolumn{1}{l}{$p_{20}$} & $-0.0279$ & \multicolumn{1}{l}{$p_{30}$} & $0.3557$ \\  
			\multicolumn{1}{l}{$p_{10}$} & $422.9$ & \multicolumn{1}{l}{$p_{11}$} & $2484$ & \multicolumn{1}{l}{$p_{21}$} & $1.374$ \\   
			\multicolumn{1}{l}{$p_{01}$} & $1213$  & \multicolumn{1}{l}{$p_{02}$} & $2911$ & \multicolumn{1}{l}{$p_{12}$} & $25.19$ \\ 
			\hline
		\end{tabular}\label{Para_model_con}
	\end{table}
	
	\subsection{Human Driver Behaviors}\label{Characterized_driver_behaviors}
	
	To describe various car-following behaviors of HDVs in mixed traffic flows, IDM is adopted to characterize a wide range of driving behaviors from real driving data, as briefly introduced below:
	\begin{equation}\label{car-followng_model}
		a = a_0 \bigg( 1 - \bigg( \frac{v}{v_d} \bigg)^{\delta_0} - \bigg( \frac{s^*(v, \Delta v)}{s} \bigg)^2 \bigg)
	\end{equation}
	with
	\begin{equation}
		s^*(v, \Delta v) = s_0 + \max \bigg( 0, T v + \frac{v \Delta v}{2 \sqrt{a_0 b_0}} \bigg)
		\nonumber
	\end{equation}
	where $v$ and $a_0$ denote the velocity and maximum acceleration respectively, $v_d$ is the desired velocity, $\delta_0$ is the acceleration exponent, $s$ and $s^*$ respectively denote the actual and desired gap distances, $s_0$ represents the minimum gap distance, $T$ is the time headway to the preceding vehicle, $\Delta v$ is the velocity difference to the preceding vehicle, and $b_0$ is the maximum deceleration.
	Then, the field-collected Next Generation Simulation (NGSIM) \cite{alexiadis2004next} dataset is applied to calibrate the adopted IDM, such that the realistic driver behavioral diversity is captured by different parameters of IDM. In order to balance the optimization performance and computational efficiency, the most dominant parameter $T$ is selected to represent different preferences of real human drivers whereas the other parameters $a_0$, $\delta_0$, $v_d$, $s_0$, and $b_0$ are set to constants after trial and error \cite{wang2022adaptive,li2023physics}. These parameters are listed in Table \ref{Para_IDM}.
	\begin{table}[htb] 
		\centering 
		\caption{IDM PARAMETERS}
		\begin{tabular}{l|l} \hline 
			IDM Parameter & Value \\ \hline
			Maximum acceleration $a_0$ & 4 \\ 
			Acceleration exponent $\delta_0$ & 4 \\ 
			Desired velocity $v_d$ & 25 \\ 
			Minimum gap distance $s_0$ & 2 \\ 
			Maximum deceleration $b_0$ & -5 \\ \hline
		\end{tabular}\label{Para_IDM}
	\end{table}
	
\begin{figure}[th]
		\centering
		\includegraphics[width=1\linewidth]{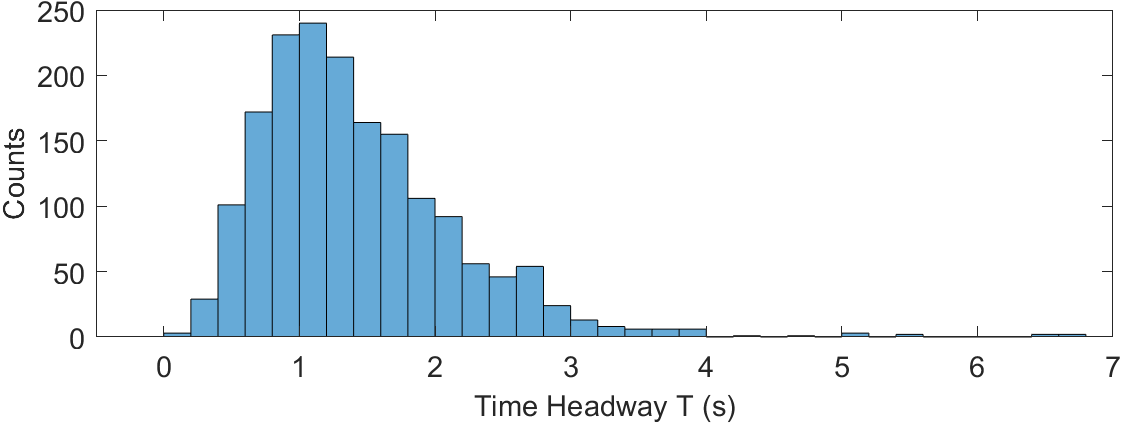}
		\caption{Distribution of time headway.}
		\label{fig_driving_behavior}
	\end{figure}

	As shown in Fig. \ref{fig_driving_behavior}, the distribution of time headway $T$, which contains $1737$ car-following cases sampled from the NGSIM dataset, is obtained by fitting the IDM parameter $T$ to the velocity trajectories. Each point in the distribution represents a certain driving style, and the range of distribution indicates the driver behavioral diversity. In this paper, the majority of driving preferences $T \in [0.5, 3]$ are considered, which includes $1626$ car-following cases for the eco-driving performance verification. 
	Moreover, to emulate the stochastic car-following behavior, a truncated normally distributed noise is added to the dynamics of IDM  under each driving style.
	
	\section{Preliminaries}
	
	\subsection{Tube-based Model Predictive Control}
	
	To keep the states $x(k)$ of the uncertain system \re{dy_act_carfollowing} close to the states $\bar x(k)$ of the nominal system \re{dy_nom_carfollowing}, the feedback controller is designed as
	\begin{equation}\label{RMPC_controller}
		u_c(k) = \bar u_c(k) + K(x(k) - \bar x(k))
	\end{equation}
	where $K \in \mathbb{R}^{1\times 2}$ is the control gain and $\bar u_c(k)$ is the control input of the nominal system. Substituting the controller \re{RMPC_controller} into the dynamics of $\tilde x(k)$ given in \re{deviation_error} yields
	\begin{equation}\label{uncertain_system}
		\tilde x(k + 1) = A_K \tilde x(k) + w(k)
	\end{equation}
	where $A_K = A + B_c K$ is stable by properly selecting the control gain $K$, e.g., solving the discrete linear quadratic regulator problem. 
	Let $\mathbb{Z}$ be a disturbance invariant set for the controlled uncertain system \re{uncertain_system}, which follows that $A_K \mathbb{Z} \oplus \mathbb{W} \subset \mathbb{Z}$.
	To reduce conservativeness, it is desirable that $\mathbb{Z}$ is as small as possible, and the corresponding minimal robust positively invariant (mRPI) set can be obtained by approximation method \cite{Rakovic2005Invariant}.
	To guarantee the satisfaction of the original constraints \re{System_constraints} under application of the controller \re{RMPC_controller}, the state and input constraints are tightened for the nominal system, that is
	\begin{align}
		\bar x(k) \in&~ \bar {\mathbb{X}}:= \mathbb{X} \ominus \mathbb{Z} \label{tight_state_cons}\\
		\bar u_c(k) \in&~ \bar {\mathbb{U}}:= \mathbb{U} \ominus K\mathbb{Z}. \label{tight_input_cons}
	\end{align}
	\begin{assumption}
		The mRPI set, i.e., $\mathbb{Z}$ is sufficiently small  such that the tightened sets $\bar {\mathbb{X}}$ and $\bar {\mathbb{U}}$ exist.
	\end{assumption}
	
	\subsection{TD3-Based Reinforcement Learning Algorithm}
	
	As the space of control input $u_c(k)$ given in \re{vehicle_model} is continuous, 
	the twin delayed deep deterministic policy gradient algorithm (TD3) \cite{fujimoto2018addressing}, which can perform over the continuous action space and reduce overestimation bias, is implemented to address the eco-driving problem.
	In the TD3 method, a policy network deterministically maps states to actions, while the minimum estimate among two Q-value networks approximates the value function.
	At each time step $t$, the TD3 agent observes a state $s_t$, and takes an action $a_t$ based on the current policy $\pi(a_t|s_t)$. Then, the environment returns a reward $r_t$ and transits to a new state $s_{t+1}$. Accordingly, the Q-value network is updated by minimizing the following loss function below.
	\begin{equation}
		L = \frac{1}{N} \sum \bigg(y_t - Q_i(s_t,a_t| \theta^{Q_i}) \bigg)^2
		\nonumber
	\end{equation}
	with
	\begin{equation}
		y_t = r_t + \gamma \min_{i=1,2} {Q '}_i \bigg(s_{t+1}, {\pi '}(s_{t+1}| \theta^{\pi '}) |\theta^{Q_i '} \bigg)
		\nonumber
	\end{equation}
	where $\pi(s | \theta^{\pi} )$, parameterized by $\theta^{\pi}$, is the policy network, $Q_i(s_t,a_t| \theta^{Q_i})$, parameterized by $\theta^{Q_i}$, is the $i$th Q-value network, $\pi '$ as the target policy network has the same structure as $\pi$ with parameter $\theta^{\pi '}$, and $Q_i '$ as the target Q-value network has the same structure as $Q_i$ with parameter $\theta^{Q_i '}$.
	The weights of target networks are updated periodically to accurately match the weights of the current networks.
	The parameters of the policy network are updated by the gradient calculated from the Q-value network as follows
	\begin{equation}
		\nabla_{\theta^{\pi}} J \approx \frac{1}{N} \sum \nabla_a Q_1(s,a| \theta^{Q_1}) \nabla_{\theta^{\pi}} \pi(s|\theta^{\pi}).
		\nonumber
	\end{equation}
	
	\section{Safe Learning-based Control for Eco-driving}\label{SafeRL}
	
	\begin{figure*}[th]
		\hspace{-2cm}
		\centering
		\includegraphics[width=0.6\linewidth]{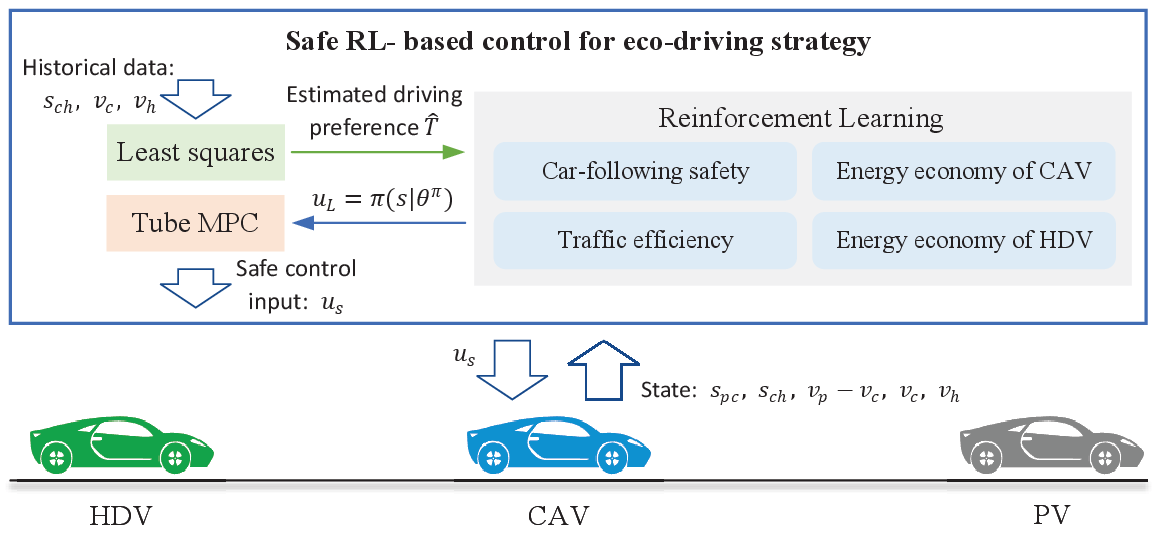}
		\caption{Illustration of eco-driving framework.}
		\label{Eco_driving_framework}
	\end{figure*}
	
	In this section, the RMPC-certified RL method for eco-driving in mixed traffic is developed, where the RMPC (i.e., tube-based MPC) ensures the safe execution of the RL policy after actions $u_L$ are selected to optimize energy consumption, as shown in Fig. \ref{Eco_driving_framework}. 
	The incorporated RMPC enables the safe learning and deployment of eco-driving policies in real-world applications, despite the presence of disturbances.
	Furthermore, the least squares method is employed to online identify the driving preference of the following HDV, and the impacts of diverse and stochastic driving behaviors of HDV on eco-driving performance are investigated.
	
	\subsection{Markov Decision Process}\label{MDP_Eco_driving}
	
	The RL problems are formalized Markov Decision Process (MDP), where an agent observes the states from the surrounding environment, takes actions, and receives rewards. Accordingly, in the eco-driving problem, the state space consists of distance gaps between vehicles and velocities, that is
	\begin{equation}\label{RL_state}
		S = \{s_{pc}, s_{ch}, v_p - v_c, v_c, v_h \}
	\end{equation}
	where $s_{pc} = s_p - s_c$ and $s_{ch} = s_c - s_h$ with $s_i$ and $v_i$, $i = \{p, c, h \}$ being defined in \re{vehicle_model}. The acceleration control input of CAV constitutes the action space, which is constrained within the same feasible region of \re{System_constraints}, that is, $u_L(k) \in [-u_{\min}, u_{\max}]$.
	However, implementing this action directly to the CAV may result in safety hazards such as inter-vehicle collisions. To handle this problem, the action provided by the RL policy is incorporated into the RMPC scheme to generate a safe control input. Moreover, in order to minimize the energy consumption of both the CAV and the HDV while guaranteeing traffic efficiency and car-following safety, the reward function includes the following four objectives.
	
	\textit{1) Energy Economy of CAV:} 
	In terms of the energy consumption model \re{Energy_model_poly}, the reward of CAV's energy economy is constructed as
	\begin{equation}\label{reward_CAV}
		r_c = - \alpha_c P_{\textrm{mot}}(v_c,u_c) \tau
	\end{equation}
	where $r_c$ is restricted to the set $r_c \in \mathbb{X}_c = [-1,1]$, $\alpha_c$ is the weight of CAV's energy economy, and $P_{\textrm{mot}}$ is the instantaneous energy consumption defined in \re{Energy_model_poly}.
	
	\textit{2) Energy Economy of HDV:} 
	Since different human drivers exert various driving behaviors, the dynamics of HDV are stochastic. This implies that the CAV cannot precisely obtain the instantaneous energy consumption of HDV according to \re{Energy_model_poly}, which poses a significant challenge for the RL training process. Hence, optimizing HDV's energy consumption requires an online estimation of HDV's acceleration. The estimated HDV's energy economy is designed in the reward function as
	\begin{equation}\label{reward_HDV}
		r_h = - \alpha_h P_{\textrm{mot}}(v_h,\hat a_h) \tau
	\end{equation}
	where $r_h$ is restricted to the set $r_h \in \mathbb{X}_h = [-1,1]$, $\alpha_h$ is the weight of HDV's energy economy, $\hat a_h$ is the estimated dynamics of HDV which will be discussed later in Section \ref{Handling_Driver_Behaviors}.
	
	\textit{3) Traffic Efficiency:}
	Within a safe range, a small time gap (TG) from the CAV to its PV indicates a higher roadway capacity \cite{wang2022adaptive,zhang2007examining}. Accordingly, the reward in terms of traffic efficiency is given by
	\begin{equation}\label{reward_traffic}
		r_t = \left\{
		\begin{array}{ll}
			- \alpha_t s_{pc},  &  \textrm{TG} \geq 2.5 \\
			0,  &  \textrm{TG} < 2.5
		\end{array}
		\right.
	\end{equation}
	with $\textrm{TG} = \frac{s_{pc}}{v_c}$
	where $\alpha_t $ is the weight of traffic efficiency and $s_{pc}$ is the distance gap defined in \re{RL_state}.
	
	\textit{4) Car-following Safety:}
	Even though the incorporated RMPC can prevent collisions between CAV and its PV, it is desirable that the RL can learn safe driving from potentially hazardous driving behaviors. 
	Time to collision (TTC) as a safety indicator implies the time span left before two vehicles collide \cite{VOGEL2003427}. Thus, the reward corresponding to the safety is designed as
	\begin{equation}\label{reward_safe}
		r_s  = \left\{
		\begin{array}{ll}
			\log(\frac{\textrm{TTC}}{4}),  & 0 \leq \textrm{TTC} \leq 4 \\
			0,  &  \textrm{else}
		\end{array}
		\right.
	\end{equation}
	where $\textrm{TTC} = - \frac{s_{pc}}{v_{p} - v_{c}}$, $\forall v_{p} < v_{c}$
	with $s_{pc}$ being the gap distance given in \re{RL_state}. Therefore, it follows from \re{reward_CAV}, \re{reward_HDV}, \re{reward_traffic}, and \re{reward_safe} that the total reward function is constructed as
	\begin{equation}
		r = r_c + r_h + r_t + r_s.
		\nonumber
	\end{equation}
	
	\subsection{Driving Behavior Estimation}\label{Handling_Driver_Behaviors}
	
	Although one parameter of IDM, i.e., the time headway $T$, is selected to characterize human driver behavior as discussed in Section \ref{Characterized_driver_behaviors}, this parameter is random and unknown prior to the CAV in real car-following scenarios. Thus, the driving preference of HDV requires to be identified online from the historical driving behaviors such that the dynamics of HDV can be further estimated. 
	To handle the diverse and stochastic driving behaviors, the historical data, including velocities $v_c$ and $v_h$ as well as distance gap $s_{ch}$, is collected, as demonstrated in Fig. \ref{Eco_driving_framework}. The IDM is then fitted to this data using the least squares method. It is noted that at the initial time of RL training, i.e., $t = 0$$s$, there is no historical data and the driving preference parameter is set to the value with a high probability, e.g., $T = 1$$s$, in terms of the distribution of $T$.
	Thus, the estimated dynamics of HDV is obtained
	\begin{equation}\label{estimated_acc_HDV}
		\hat a_{h} = f_{\textrm{IDM}}(\hat T, v_h, s_{ch})
	\end{equation}
	where $\hat T$ is the estimated time headway, i.e., driving preference, which is obtained using the least squares method, $v_h$ and $s_{ch}$ are defined in \re{vehicle_model} and \re{RL_state} respectively, and $f_{\textrm{IDM}}$ is the abbreviation for IDM defined in \re{car-followng_model}. 
	
	\subsection{Model Predictive Safety Certification}
	
	In this study, the RL algorithm is implemented to solve the eco-driving problem, in which a TD3 agent is designed to learn an optimal energy efficiency policy by maximizing the discounted total reward as discussed in Section \ref{MDP_Eco_driving}. 
	Since the TD3 agent needs to learn a safe eco-driving policy from hazardous repeated collisions, its action, that is, $u_L(k) = \pi(s_{t = k}|\theta^{\pi})$ cannot be directly applied to the CAV without any safety certification during operation of the real system; otherwise, undesirable accidents might happen. 
	Furthermore, there is no guarantee that the trained policy can avoid collisions at all times when applied to real systems.
	
	To address this problem, before an action $u_L(k)$ is applied to the CAV, it needs to be integrated into the RMPC scheme to generate a safe control input $u_s(k)$ as illustrated in Fig. \ref{Eco_driving_framework}. Inspired by \cite{Wabersich2018Linear,WABERSICH2021109597}, the idea is to drive the first input of the control sequence closer to the action of the TD3 agent, while respecting all safety constraints. Due to the RMPC only addressing the constraint requirements under disturbances, without considering the issues of energy consumption and driving behaviors, it is straightforward to formulate the RMPC problem as follows
	\begin{subequations}\label{Cost_MPCRL}
		\begin{align}
			\min_{\bar x_{(0|k)}, \bar u_{c(i|k)}} \sum_{i=0}^{N-1} &~ l(\bar x(i), \bar u_c(i)) + V_f(\bar x(N))
			+ V_l(u_L, \bar u_{0}) \label{RMPC_RL}\\
			\textrm{s.t.}~ \bar x_{(i+1|k)} =&~  A \bar x_{(i|k)} + B_c \bar u_{c(i|k)} + B \bar   a_{p(i|k)}\label{RMPC_RL_inq_cons} \\
			\bar x_{(i|k)} \in&~  \bar {\mathbb{X}}, ~ i = 0,1,..., N-1 \label{RMPC_RL_state_cons}\\
			\bar u_{c(i|k)} \in&~  \bar {\mathbb{U}}, ~ i = 0,1,..., N-1 \label{RMPC_RL_input_cons}\\
			\bar x_{(N|k)} \in&~  {\mathbb{X}}_f \label{RMPC_RL_ter_cons}\\
			x(k) \in&~ \bar x_{(0|k)} \oplus  \mathbb{Z} \label{RMPC_RL_init_cons}
		\end{align}
	\end{subequations}
	with
	\begin{align}
		l(\bar x(i), \bar u_c(i)) =&~ \bar x_{(i|k)}^T Q \bar x_{(i|k)} + \bar u_{c(i|k)}^T R \bar u_{c(i|k)}\nonumber\\
		V_f(\bar x(N)) =&~ \bar x^T_{(N|k)} P \bar x_{(N|k)}\nonumber\\
		V_l(u_L, \bar u_0) =&~  ( \bar u_{c(0|k)} - u_L(k))^T R_l (\bar u_{c(0|k)} - u_L(k))
		\nonumber
	\end{align}
	where  $N$ is the predictive horizon, $\bar x_{(i|k)}$ and $\bar u_{c(i|k)}$ denote the predicted $i$th nominal system state and input at time step $k$, respectively, $x(k)$ is the measured state of actual system, $l(\cdot)$ is the stage cost, $V_f(\cdot)$ is the terminal cost, $V_l(\cdot)$ is the cost that minimizes the difference of the first input of the control sequence $\bar u_{c(0|k)}$ and the action of TD3 $u_L(k)$, weighting matrices $Q$, $R$, $P$, and $R_l$ are symmetric and positive, the equality constraint \re{RMPC_RL_inq_cons} is given in \re{dy_nom_carfollowing}, $\bar {\mathbb{X}}$ and $\bar {\mathbb{U}}$ are the tightened state and input constraints given in \re{tight_state_cons} and \re{tight_input_cons}, respectively, ${\mathbb{X}}_f$ is the terminal set, and $\mathbb{Z}$ is the mRPI.
	The solution of \re{Cost_MPCRL} yields the optimal control sequence $\bar u_c^{*}(k) = \{ \bar u_{c(0|k)}^{*} , \bar u_{c(1|k)}^{*}, ..., \bar u_{c(N-1|k)}^{*}\}$ and the associated optimal state sequence $\bar x^{*}(k) = \{ \bar x_{(0|k)}^{*}, \bar x_{(1|k)}^{*}, ..., \bar x_{(N|k)}^{*} \}$. Consequently, the controller resulting from addressing the optimization problem \re{Cost_MPCRL} is given by
	\begin{equation}\label{safe_control_law}
		u_s(k) = \bar u_{c(0|k)}^{*} + K(x(k) - \bar x_{(0|k)}^{*})
	\end{equation}
	where $u_s(k)$ is the safe control input applied to the CAV, that is, $u_c(k) = u_s(k)$, and $K$ is the control gain given in \re{uncertain_system}.
	
	Let \re{Cost_MPCRL} be feasible and $x(0) \in \mathbb{X}$. The condition \re{RMPC_RL_init_cons} implies that $\tilde{x}(k) = x(k) - \bar x(k) \in \mathbb{Z}$. Furthermore, it is from \re{uncertain_system} that $A_K$ stabilizes the error dynamics, which implies $\tilde x(k+1) = A_K \tilde{x}(k) + w(k) \in A_K \mathbb{Z} \oplus \mathbb{W} \subset \mathbb{Z}$. Thus, $x(k+1) = \bar x(k+1) + \tilde{x}(k+1) \in \bar x(k+1) \oplus \mathbb{Z} $. According to the condition \re{RMPC_RL_state_cons} and the tightened state constraint \re{tight_state_cons}, $x(k+1) \in \bar {\mathbb{X}} \oplus \mathbb{Z} \subset \mathbb{X}$. 
	Moreover, the condition \re{RMPC_RL_input_cons} indicates that $\bar u_c^* (k) \in \bar {\mathbb{U}}$, and it is from the condition \re{RMPC_RL_init_cons}, the tightened input constraint \re{tight_input_cons}, and the control law \re{safe_control_law} that $u_s(k) = \bar u_c^* (k) + K \tilde{x}(k) \in \bar {\mathbb{U}} \oplus K \mathbb{Z} \subset \mathbb{U}$.
	Therefore, due to the facts $x(k+1) \in \mathbb{X}$ and $u_s(k) \in \mathbb{U}$, the state and input constraints imposed on the actual system are not violated.
	
	\section{Simulation Results}\label{Simulation}
	\subsection{Simulation Setup}
	In this section, simulation studies are conducted to verify the safety and eco-driving performance of the proposed method. Additionally, we investigate the impacts of diverse and stochastic driving behaviors of HDV on eco-driving performance. 
	
	During the training process, the number of training episodes is set to $5000$, with each episode consisting of $300$ steps. The PV follows a designed velocity profile, which is sampled from the NGSIM dataset to emulate the real-world driving scenario. The dynamics of HDV are updated by the randomly parameterized IDM in each episode. Specifically, as illustrated in Fig. \ref{fig_driving_behavior}, the diversity in human drivers is characterized by the range of distribution of $T$. Thus, by uniformly interpolating $100$ data points in the range of $T \in [0.5, 3]$, $100$ different driving preferences are obtained. In each episode, one of a hundred driving preferences is randomly sampled to update the dynamics of HDV, which is unknown to the CAV. Furthermore, the noise is added to the dynamics of HDV to represent the stochastic behaviors.
	The initial conditions of vehicles are set to $s_{pc}(0) = 20$$m$, $s_{ch}(0) = 20$$m$, $v_p(0) - v_c(0) = 1.6416$$m/s$, $v_c(0) = 8.5$$m/s$, and $v_h(0) = 8$$m/s$ with the initial estimated driving preference parameter being $\hat T_0 = 1$$s$. The hyper-parameter configurations and network structure of the TD3 algorithm are listed in Table \ref{Para_TD3}. The state, action, and reward in MDP are discussed in Section \ref{MDP_Eco_driving}.
	
	\begin{table}[htbp] 
		\centering 
		\caption{Hyper-parameter Settings and Network Structure of TD3}
		\begin{tabular}{l|l|l|l|l} \hline 
			\multicolumn{3}{l}{Hyperparameters} & \multicolumn{2}{l}{Value} \\ \hline
			\multicolumn{3}{l}{Replay buffer capacity} & \multicolumn{2}{l}{$20000$} \\ 
			\multicolumn{3}{l}{Discount factor} & \multicolumn{2}{l}{$0.9$} \\ 
			\multicolumn{3}{l}{Sample batches size} & \multicolumn{2}{l}{$16$} \\ 
			\multicolumn{3}{l}{Learning rate of policy network} & \multicolumn{2}{l}{$0.00001$} \\ 
			\multicolumn{3}{l}{Learning rate of Q-value network} & \multicolumn{2}{l}{$0.00002$} \\ 
			\multicolumn{3}{l}{Target network update rate} & \multicolumn{2}{l}{$0.005$} \\
			\multicolumn{3}{l}{Frequency of delayed policy updates} & \multicolumn{2}{l}{$2$} \\
			\multicolumn{3}{l}{Gaussian exploration noise} & \multicolumn{2}{l}{$0.9992^{i}$} \\ 
			\multicolumn{3}{l}{Q-value network update noise} & \multicolumn{2}{l}{$0.1$} \\
			\multicolumn{3}{l}{Range to clip target policy noise} & \multicolumn{2}{l}{$0.1$} \\ 
			\hline
			\multicolumn{5}{l}{Network Structure}   \\ 
			\hline
			\multicolumn{1}{l}{ } & \multicolumn{2}{l}{Policy network} & \multicolumn{2}{l}{Q-value network}  \\ 
			\multicolumn{1}{l}{ } & \multicolumn{1}{l}{Nodes} & \multicolumn{1}{l}{Activation} & \multicolumn{1}{l}{Nodes}& \multicolumn{1}{l}{Activation}  \\ 
			\multicolumn{1}{l}{Input Layer } & \multicolumn{1}{l}{$5$} & \multicolumn{1}{l}{} & \multicolumn{1}{l}{$6$}& \multicolumn{1}{l}{}  \\ 
			\multicolumn{1}{l}{Hidden Layer } & \multicolumn{1}{l}{$256$} & \multicolumn{1}{l}{ReLU} & \multicolumn{1}{l}{$256$}& \multicolumn{1}{l}{ReLU}  \\ 
			\multicolumn{1}{l}{Hidden Layer } & \multicolumn{1}{l}{$128$} & \multicolumn{1}{l}{ReLU} & \multicolumn{1}{l}{$128$}& \multicolumn{1}{l}{ReLU}  \\ 
			\multicolumn{1}{l}{Output Layer } & \multicolumn{1}{l}{$1$} & \multicolumn{1}{l}{Tanh} & \multicolumn{1}{l}{$1$}& \multicolumn{1}{l}{}  \\ 
			\hline
			\multicolumn{5}{l}{$i$: the $i$th episode}
		\end{tabular}\label{Para_TD3}
	\end{table}
	
	To simulate the real situation, the disturbances resulted from prediction uncertainty and measurement errors are taken into account. Specifically, a truncated normally distributed noise is added to the acceleration of PV, that is, $\Delta a_p \sim \mathcal{N}(0,\sigma^2)$ with the truncated interval $[-n_p, n_p]$. Thus, the predicted dynamics of PV required in \re{RMPC_RL_inq_cons} is constructed as 
	\begin{equation}
		\bar a_p = a_p + \Delta a_p
		\nonumber
	\end{equation}
	where $a_p$ is obtained from the velocity profile in the NGSIM dataset. Additionally, the measured position and velocity of PV are contaminated by uniformly distributed noise, that is
	\begin{equation}
		x_p(k+1) = A (x_p(k) + \Delta_p) + B a_p(k)
		\nonumber
	\end{equation}
	where $\Delta_p = [\Delta_s, \Delta_v]^T$ with $\Delta_s \leq b_s$ and $\Delta_v \leq b_v$. It can be verified that the disturbance resulted from prediction uncertainty and measurement errors is bounded by $c_w$. For the stochastic behaviors of the HDV, its dynamics are contaminated by a normally distributed noise $\Delta a_h \sim \mathcal{N}(0,\sigma^2)$ with the truncated interval $[-n_h, n_h]$. Thus, the stochastic dynamics of HDV is described by 
	\begin{equation}
		x_h(k+1) = Ax_h(k) + B (1 + \Delta a_h) a_h(k).
		\nonumber
	\end{equation}
	Relevant parameters in the simulation are listed in Table \ref{Sim_Para}.
	
	\begin{table}[htbp] 
		\centering 
		\caption{The Parameters Values Used in Simulation Study}
		\begin{tabular}{l|l|l|l|l|l} \hline 
			\multicolumn{1}{l}{Parameter} & Value& \multicolumn{1}{l}{Parameter} & Value & \multicolumn{1}{l}{Parameter} & Value \\ \hline
			\multicolumn{1}{l}{$\tau$} & $0.5$ & \multicolumn{1}{l}{$h$} & $0.5$ & \multicolumn{1}{l}{$x_{1,\min}$} & $2$ \\
			\multicolumn{1}{l}{$x_{2,\min}$} & $5$ & \multicolumn{1}{l}{$x_{2,\max}$} & 5 & \multicolumn{1}{l}{$u_{\min}$} & $3$ \\
			\multicolumn{1}{l}{$u_{\max}$} & $3$ & \multicolumn{1}{l}{$\alpha_c$} & $1/30000$ & \multicolumn{1}{l}{$\alpha_h$} & $1/30000$ \\
			\multicolumn{1}{l}{$\alpha_t$} & $1/25$ & \multicolumn{1}{l}{$Q$} & $\textrm{diag}(1,1)$ & \multicolumn{1}{l}{$R$} & $1$ \\
			\multicolumn{1}{l}{$R_l$} & $50$ & \multicolumn{1}{l}{$N$} & $50$ & \multicolumn{1}{l}{$c_w$} & $0.3$\\
			\multicolumn{1}{l}{$\sigma$} & $0.1$ & \multicolumn{1}{l}{$n_h$} & $0.05$ & \multicolumn{1}{l}{$n_p$} & $0.2$\\
			\multicolumn{1}{l}{$b_s$} & $0.1$ & \multicolumn{1}{l}{$b_v$} & $0.2$ & \multicolumn{1}{l}{} &  
			\\ \hline
		\end{tabular}\label{Sim_Para}
	\end{table}
	
	\subsection{Safety and Economy Performance }\label{Sim_A}
	
	\begin{figure}[tbp]
		\centering
		\includegraphics[width=0.85\linewidth]{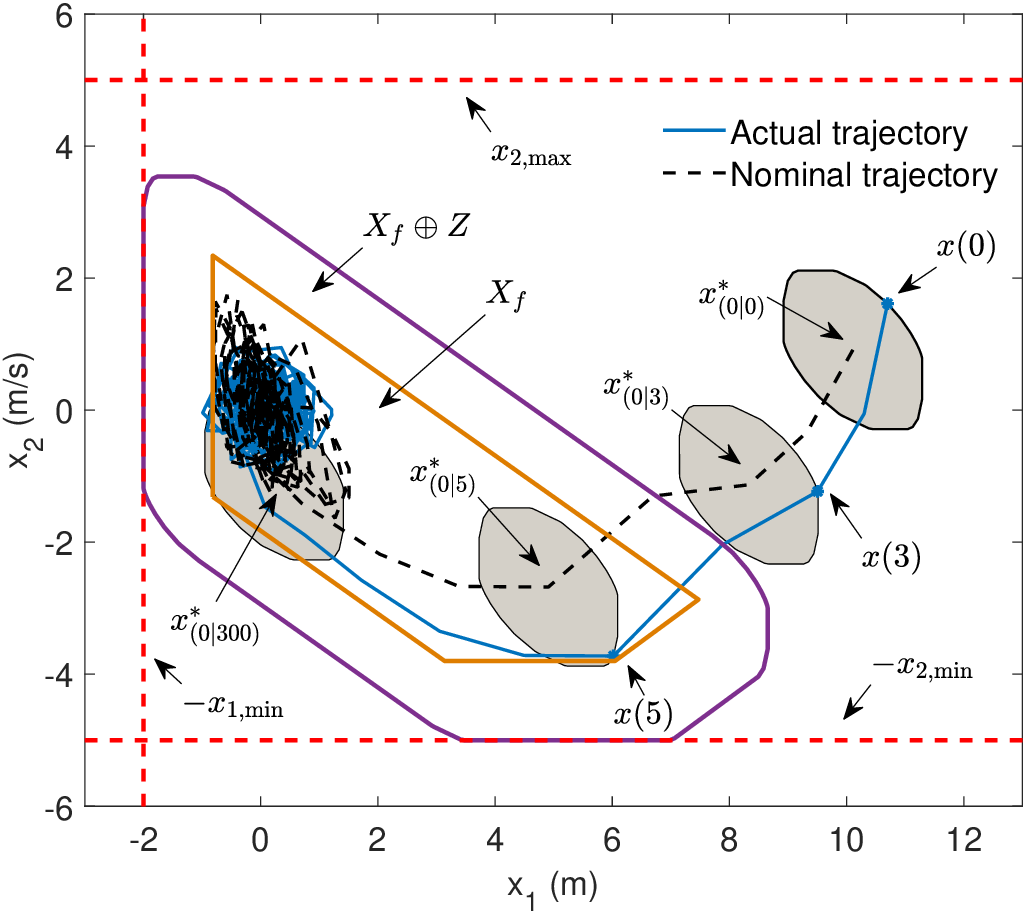}
		\caption{System trajectories using the RMPC controller.}
		\label{tube}
	\end{figure}
	
	\begin{figure}[!h]
		\centering
		\subfigure[]{
			\begin{minipage}[t]{1\linewidth}
				\includegraphics[width=0.8\linewidth]{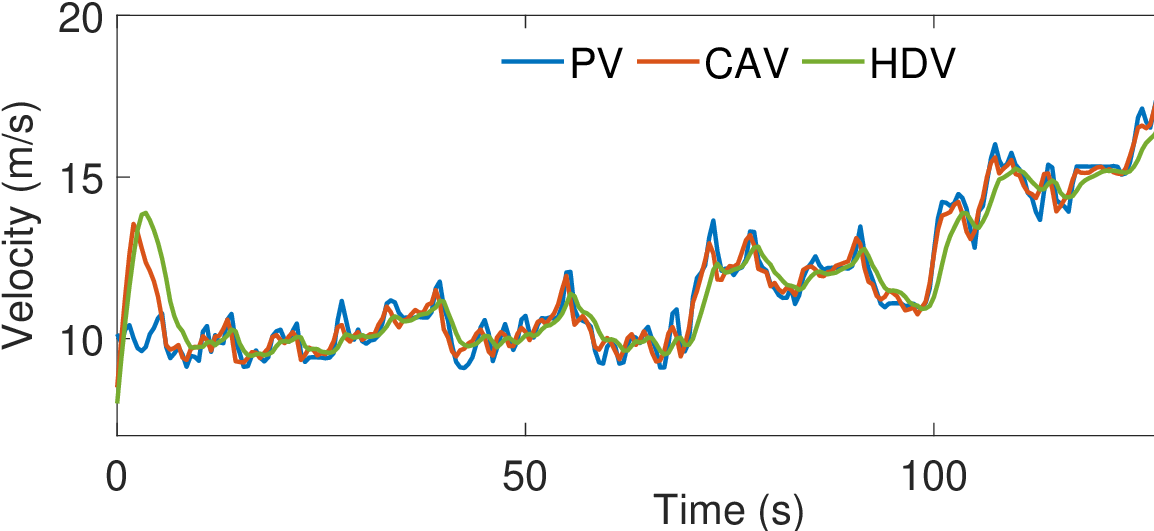}
			\end{minipage}
			\label{velocity_RMPC}
		}

		\subfigure[]{
			\begin{minipage}[t]{1\linewidth}
				\includegraphics[width=0.8\linewidth]{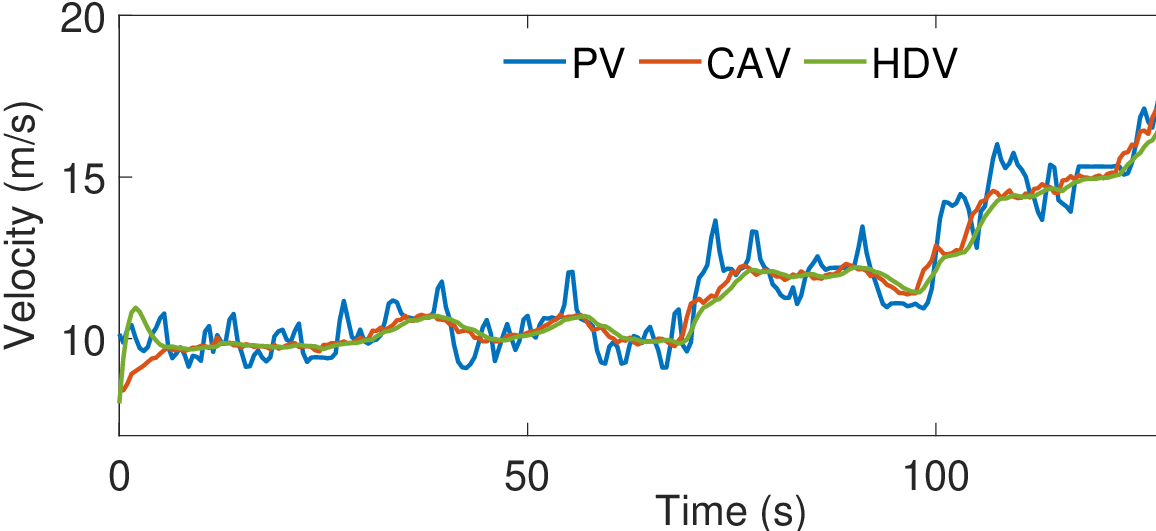}
			\end{minipage}
			\label{velocity_RMPCRL}
		}  
		
		\subfigure[]{
			\begin{minipage}[t]{1\linewidth}
				\includegraphics[width=0.8\linewidth]{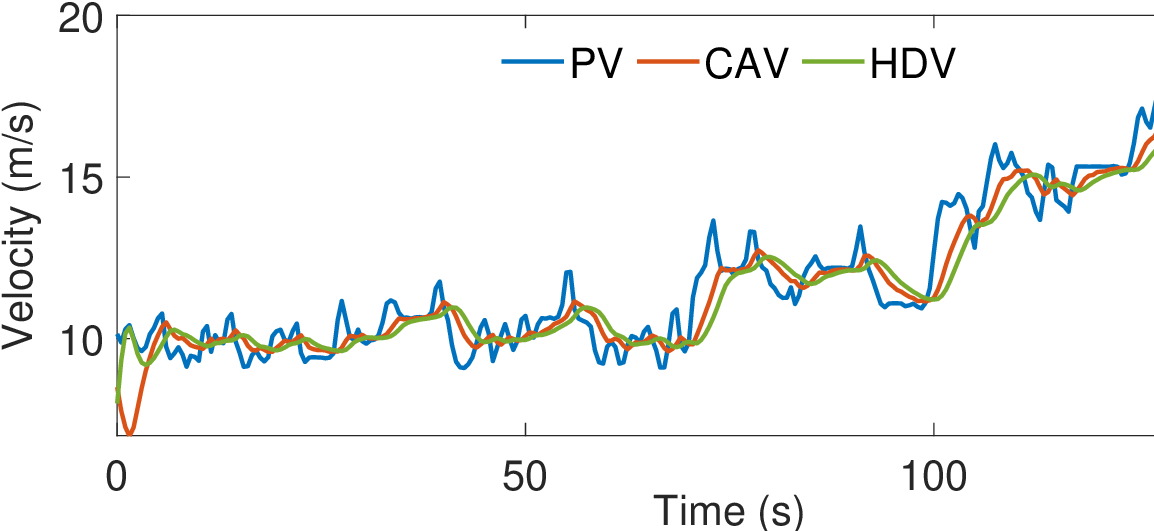}
			\end{minipage}
			\label{velocity_RL}
		}
		\caption{Velocity profiles of vehicles using different algorithms. (a) RMPC. (b) RMPC-certified RL. (c) RL.}
		\label{velocity_profiles}
	\end{figure}
	
\begin{figure}[t]
		\centering
		\includegraphics[width=0.85\linewidth]{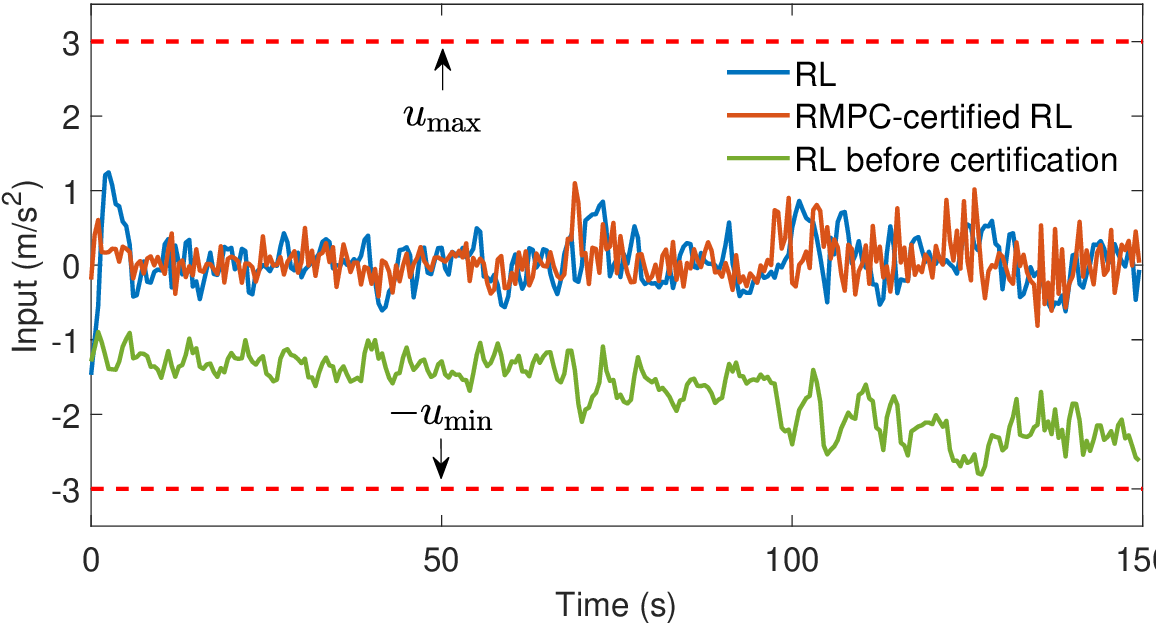}
		\caption{Control inputs comparison between RL and RMPC-certified RL algorithms.}
		\label{Compare_input}
	\end{figure}

	To demonstrate the safety and eco-driving performance of the proposed strategy, comparative simulation studies are performed using the presented RMPC-certified RL algorithm, RMPC technique, and RL algorithm for the mixed traffic scenario. 
	The hyper-parameter configurations and network structure of the RL algorithm are the same as that of the RMPC-certified RL algorithm except for the learning rate of the Q-value network being $0.00005$.
	The parameters of the RMPC are identical to that of the RMPC-certified RL with the cost $V_l(\cdot)$ in \re{Cost_MPCRL} being removed.
	The initial conditions for the test are identical to that of the training process, except for $s_{pc}(0) = 15$$m$. Comparative results of a car-following scenario with the time headway $T = 1.2$$s$ for the HDV using different algorithms are shown in Figs. \ref{tube}--\ref{Compare_input}.
	
	\begin{figure}[th]
		\centering
		\includegraphics[width=0.88\linewidth]{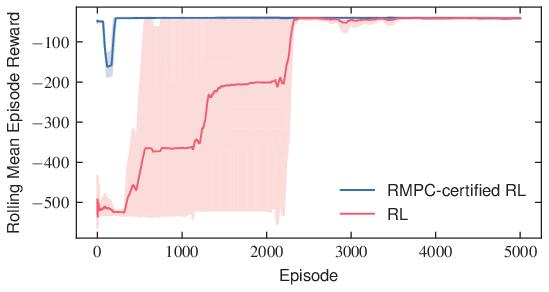}
		\caption{Training results comparison between RL and RMPC-certified RL.}
		\label{Compare_reward}
	\end{figure}
	
	In Fig. \ref{tube}, the blue solid line denotes the actual trajectory $\{x(i)\}$ of the car-following system while the black dash line represents the optimal state sequence $\{x^{*}_{(0|k)}\}$ of the corresponding nominal system. Using the RMPC controller, both trajectories lie within the feasible region constructed by the constraints (red dash lines), which implies that the constraint requirements are not violated despite the presence of disturbances. 
	Similarly, by solving the problem \re{Cost_MPCRL}, the RMPC-certified RL algorithm guarantees the state and input constraints not only during the training process but also when the trained eco-driving policy is applied to real systems. 
	By contrast, with different random seeds, the RL algorithm learns a safe eco-driving strategy from $231$, $147$, and $2048$ collisions with its PV respectively. 
	In addition, there is no guarantee that this trained policy can achieve safe driving at all times when applied to real systems. Thus, the presented RMPC-certified RL algorithm outperforms the RL algorithm in terms of safety.
	
	Since the RMPC method only minimizes the states of control input, i.e., the acceleration of CAV while ensuring the constraint requirements, without considering the energy consumption \re{Energy_model_poly} and driving behaviors of HDV, it cannot efficiently avoid inefficient maneuvers including unnecessary acceleration and braking, unlike RMPC-certified RL and RL algorithms. 
	This can be verified from Fig. \ref{velocity_profiles} that using the RMPC controller (Fig. \ref{velocity_RMPC}), the velocities of both the CAV and the HDV tend to closely follow the velocity of PV,  which implies higher energy consumption. 
	In comparison, the velocities of vehicles are relatively smooth using both RMPC-certified RL and RL algorithms (Figs. \ref{velocity_RMPCRL}--\ref{velocity_RL}), which indicates smaller  magnitudes of accelerations, leading to lower energy consumption. 
	
	Moreover, when the optimal solution found by the RL algorithm lies within the feasible region of the RMPC problem \re{Cost_MPCRL}, the optimal solution obtained by the RMPC-certified RL algorithm can closely follow the solution found by the RL algorithm, which implies that the similar inefficient maneuvers are avoided to achieve eco-driving. Although the TD3 agent in RL and RMPC-certified RL algorithms learns different policies (blue and green lines respectively seen in Fig. \ref{Compare_input}), incorporating the RL policy learned by the presented method into the RMPC scheme results in the safe input (brown line) that closely resembles the well-trained RL policy (blue line) without violating the input constraints.
	This can be further verified from Fig. \ref{Compare_reward} that the maximum rewards found by both RMPC-certified RL and RL algorithms tend toward consistency, which indicates that the
	learned optimal policies are similar. 
	It is noted that the reward obtained using the RL algorithm starts to rise from around $-500$ since the penalty is set to $-500$ when a collision occurs.
	
	\begin{table}[tb] 
		\centering 
		\caption{Energy Consumption Comparison Using Different Algorithms}
		\begin{tabular}{l|l|l|l} \hline 
			\multicolumn{1}{l}{} & \multicolumn{3}{c}{Holistic energy consumption}   \\ \hline
			\multicolumn{1}{l}{Algorithms} & \multicolumn{1}{l}{Least}& \multicolumn{1}{l}{Most} & \multicolumn{1}{l}{Mean}
			\\ \cline{2-4}
			\multicolumn{1}{l}{RMPC} & \multicolumn{1}{l}{$1118.82$} & \multicolumn{1}{l}{$1229.08$} & \multicolumn{1}{l}{$1157.14$} \\
			\multicolumn{1}{l}{RMPC-certified RL} & \multicolumn{1}{l}{$1013.85$} &\multicolumn{1}{l}{$1070.69$} & \multicolumn{1}{l}{$1031.30$} \\
			\multicolumn{1}{l}{RL} & \multicolumn{1}{l}{$1023.68$} &\multicolumn{1}{l}{$1062.39$} & \multicolumn{1}{l}{$1040.04$}  \\
			\hline 
			\multicolumn{1}{l}{Unit: kJ/km}
		\end{tabular}\label{Table_EnergyCompare}
	\end{table}
	
	\begin{figure}[tp]
		\centering
		\includegraphics[height=0.38\linewidth]{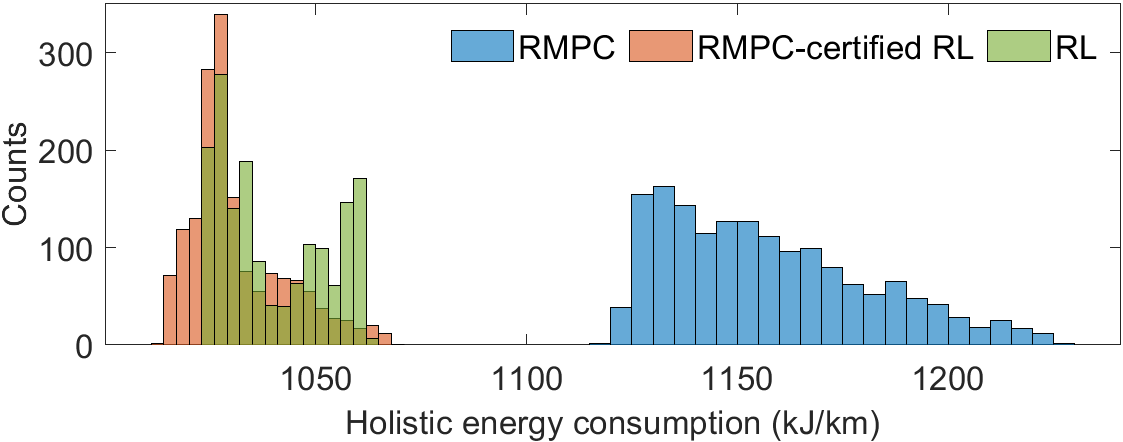}
		\caption{Holistic energy consumption distribution using different algorithms.}
		\label{EnergyCompare_distribution}
	\end{figure}
	
	To evaluate the economy performance, $1626$ car-following cases from the NGSIM dataset covering the range of time headway $T \in [0.5,3]$ are tested using different algorithms. Table \ref{Table_EnergyCompare} provides a detailed comparison of the holistic energy consumption of both the CAV and the HDV. 
    With the RMPC-certified RL algorithm, the average holistic energy consumption drops from $1157.14$$kJ/km$ to $1031.30$$kJ/km$ in comparison to the safe-oriented RMPC controller, indicating the average improvement of $10.88\%$ in energy efficiency.
    The economy performances based on the RMPC-certified RL and RL algorithms are close, resulting in average holistic energy consumptions of $1031.30$$kJ/km$ and $1040.04$$kJ/km$, respectively.
    Similar results can be found in the holistic energy consumption distribution, as shown in Fig. \ref{EnergyCompare_distribution}, where the distributions obtained by using the RMPC-certified RL and RL algorithms are highly overlapping, whereas the RMPC controller results in a distribution of higher energy consumption.
	Therefore, the presented RMPC-certified RL algorithm can achieve similar economy performance as the RL algorithm while retaining the safety characteristics of the RMPC technique, which has advantages over both RL and RMPC algorithms.

	\subsection{Impacts of Driving Behaviors on Eco-driving Performance }

	In addition, the impacts of driving behaviors of the following HDV on eco-driving performance are taken into account. 
	In \textit{Scenario A}, a PV is directly followed by the HDV.
	In \textit{Scenario B}, a CAV is inserted between the PV and the HDV, where only the energy consumption of the CAV is optimized, and the HDV directly follows the CAV.
	In \textit{Scenario C}, the platoon structure is the same as that in \textit{Scenario B}, but the energy efficiencies of both the CAV and the HDV are optimized as discussed in \ref{Sim_A}. 
        In the proposed framework, it is flexible to design RL algorithms tailored for different eco-driving problems.
    Accordingly, in \textit{Scenario B}, the HDV-related states and reward in MDP are removed, that is, $S = \{s_{pc}, v_p - v_c, v_c \}$ and $r = r_c + r_t + r_s$. 
	The configurations and parameters are listed in Tables \ref{Para_TD3} and \ref{Sim_Para}. 

        \begin{figure}[bp]
	\centering
		\includegraphics[height=0.38\linewidth]{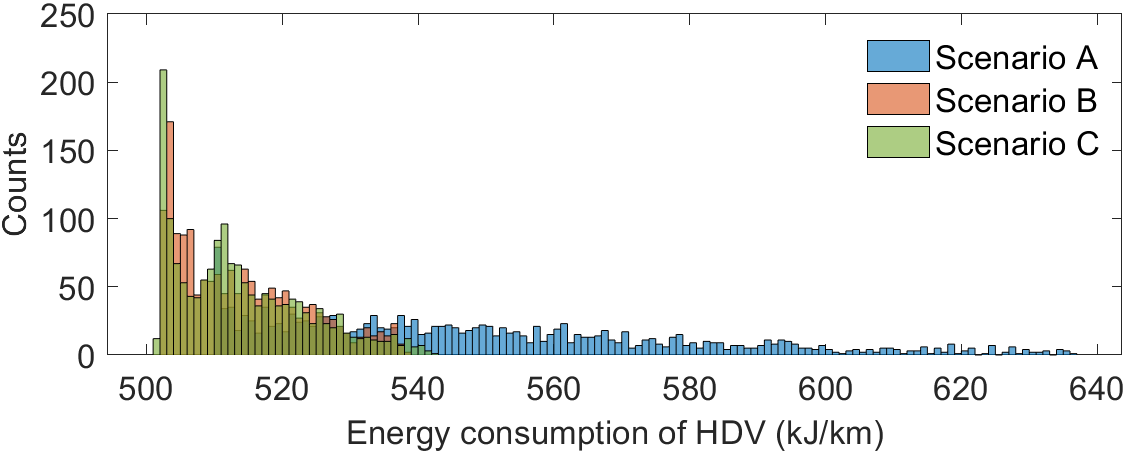}
		\caption{Energy consumption distribution of HDV in three scenarios.}
		\label{EnergyCompare_distribution_HDV}
	\end{figure}
 
	\begin{figure}[bp]
		\centering
		\includegraphics[height=0.37\linewidth]{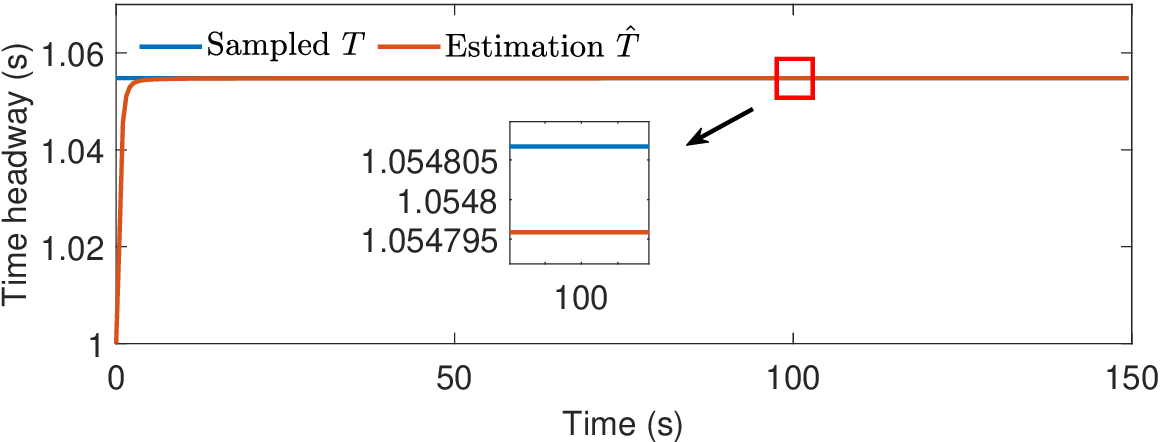}
		\caption{A random sample of car-following case.}
		\label{T_esitmation}
        \end{figure}

    As shown in Fig. \ref{EnergyCompare_distribution_HDV} where $1626$ driving preferences sampled from NGSIM dataset are tested in the above-mentioned three scenarios.
	By comparing the scenarios A and B, it can be observed that without optimizing the energy consumption of the HDV, its energy efficiency is positively affected by inserting an eco-driving CAV in the platoon. 
	This is because the motion of HDV simulated by IDM tends to follow that of its preceding vehicles (PV and CAV respectively), and thus when the preceding CAV performs eco-driving with a smooth velocity, the following HDV exhibits a similar behavior.
	Furthermore, the distributions of energy consumption in scenarios B and C highly overlap, which implies that, compared with only optimizing the energy consumption of the CAV, optimizing that of both the CAV and the HDV does not result in a significant improvement in HDV's energy efficiency. 
	Similar results can also be obtained in terms of holistic energy consumption, where 
	the average energy consumptions for the platoon in scenarios B and C are respectively $1029.30$$kJ/km$ and $1031.30$$kJ/km$.
	This implies that under the positive influence of the CAV, even if the driving preference of the HDV is estimated with a small estimation error (Fig. \ref{T_esitmation}), there is not a significant improvement in energy efficiency when considering the energy consumption of the HDV with diverse and stochastic driving behaviors.

	\section{Conclusion}\label{Conclusion}
	This paper presents a safe learning-based eco-driving framework for mixed traffic flows, which enables the safe learning and deployment of complex eco-driving policies in real-world applications.
	Specifically, the RL algorithm, which optimizes energy efficiency in intricate environments, is integrated into the optimization problem solved by the MPC scheme subject to safety constraints. 
	Tube-based enhanced MPC guarantees the satisfaction of system constraints despite the presence of disturbances, thereby improving the control robustness.
	Simulation results show that the presented RMPC-certified RL algorithm is capable of achieving similar economy performance as the RL algorithm while retaining the safety characteristics of the RMPC technique. 
	In addition, the insertion of an eco-driving CAV ahead of the HDV positively impacts the maneuvers of the HDV, leading to a significant reduction in holistic energy consumption of the mixed traffic flow.
	
	\bibliographystyle{IEEEtran}
	\bibliography{References/Ref_Eco_20231022}

\end{document}